\PassOptionsToPackage{unicode}{hyperref}
\PassOptionsToPackage{hyphens}{url}
\PassOptionsToPackage{dvipsnames,svgnames,x11names}{xcolor}
\documentclass[
]{article}
\usepackage{amsmath,amssymb}
\usepackage{lmodern}
\usepackage{iftex}
\ifPDFTeX
  \usepackage[T1]{fontenc}
  \usepackage[utf8]{inputenc}
  \usepackage{textcomp} 
\else 
  \usepackage{unicode-math}
  \defaultfontfeatures{Scale=MatchLowercase}
  \defaultfontfeatures[\rmfamily]{Ligatures=TeX,Scale=1}
\fi
\IfFileExists{upquote.sty}{\usepackage{upquote}}{}
\IfFileExists{microtype.sty}{
  \usepackage[]{microtype}
  \UseMicrotypeSet[protrusion]{basicmath} 
}{}
\makeatletter
\@ifundefined{KOMAClassName}{
  \IfFileExists{parskip.sty}{%
    \usepackage{parskip}
  }{
    \setlength{\parindent}{0pt}
    \setlength{\parskip}{6pt plus 2pt minus 1pt}}
}{
  \KOMAoptions{parskip=half}}
\makeatother
\usepackage{xcolor}
\usepackage{color}
\usepackage{fancyvrb}

\DefineVerbatimEnvironment{Highlighting}{Verbatim}{commandchars=\\\{\}}
\newenvironment{Shaded}{}{}

\newcommand{\CommentTok}[1]{\textcolor[rgb]{0.38,0.63,0.69}{\textit{#1}}}

\newcommand{\DecValTok}[1]{\textcolor[rgb]{0.25,0.63,0.44}{#1}}

\newcommand{\NormalTok}[1]{#1}
\newcommand{\OperatorTok}[1]{\textcolor[rgb]{0.40,0.40,0.40}{#1}}

\newcommand{\StringTok}[1]{\textcolor[rgb]{0.25,0.44,0.63}{#1}}

\usepackage{longtable,booktabs,array}
\usepackage{calc} 
\usepackage{etoolbox}
\makeatletter
\patchcmd\longtable{\par}{\if@noskipsec\mbox{}\fi\par}{}{}
\makeatother
\IfFileExists{footnotehyper.sty}{\usepackage{footnotehyper}}{\usepackage{footnote}}
\makesavenoteenv{longtable}
\newlength{\cslhangindent}
\newlength{\csllabelwidth}
\newlength{\cslentryspacingunit} 
\newenvironment{CSLReferences}[2] 
 {
  \setlength{\parindent}{0pt}
  \ifodd #1
  \let\oldpar\par
  \def\par{\hangindent=\cslhangindent\oldpar}
  \fi
  \setlength{\parskip}{#2\cslentryspacingunit}
 }%
 {}
\usepackage{calc}

\ifLuaTeX
\usepackage[bidi=basic]{babel}
\else
\usepackage[bidi=default]{babel}
\fi
\babelprovide[main,import]{american}

\def\languageshorthands#1{}
\ifLuaTeX
  \usepackage{selnolig}  
\fi
\IfFileExists{bookmark.sty}{\usepackage{bookmark}}{\usepackage{hyperref}}
\IfFileExists{xurl.sty}{\usepackage{xurl}}{} 
\hypersetup{
  pdftitle={Learning from Crowds with Crowd-Kit},
  pdfauthor={Dmitry Ustalov, Nikita Pavlichenko, Boris Tseitlin},
  pdflang={en-US},
  colorlinks=true,
  linkcolor={Maroon},
  filecolor={Maroon},
  citecolor={Blue},
  urlcolor={Blue},
  pdfcreator={LaTeX via pandoc}}

\title{Learning from Crowds with Crowd-Kit}

\usepackage[affil-it]{authblk}
\usepackage{orcidlink}
\author[1%
  \ensuremath\mathparagraph]{Dmitry Ustalov%
    \,\orcidlink{0000-0002-9979-2188}\,%
    }
\author[2%
  ]{Nikita Pavlichenko%
    \,\orcidlink{0000-0002-7330-393X}\,%
    }
\author[3%
  ]{Boris Tseitlin%
    \,\orcidlink{0000-0001-8553-4260}\,%
    }

\affil[1]{JetBrains, Serbia}
\affil[2]{JetBrains, Germany}
\affil[3]{Planet Farms, Portugal}
\affil[$\mathparagraph$]{Corresponding author}
\date{24 September 2023}

\begin{document}
\maketitle

\hypertarget{summary}{%
\section{Summary}\label{summary}}

This paper presents Crowd-Kit, a general-purpose computational quality
control toolkit for crowdsourcing. Crowd-Kit provides efficient and
convenient implementations of popular quality control algorithms in
Python, including methods for truth inference, deep learning from
crowds, and data quality estimation. Our toolkit supports multiple
modalities of answers and provides dataset loaders and example notebooks
for faster prototyping. We extensively evaluated our toolkit on several
datasets of different natures, enabling benchmarking computational
quality control methods in a uniform, systematic, and reproducible way
using the same codebase. We release our code and data under the Apache
License 2.0 at \url{https://github.com/Toloka/crowd-kit}.

\hypertarget{statement-of-need}{%
\section{Statement of need}\label{statement-of-need}}

A traditional approach to quality control in crowdsourcing builds upon
various organizational means, such as careful task design,
decomposition, and preparing golden tasks
(\protect\hyperlink{ref-Zhdanovskaya:23}{Zhdanovskaya et al., 2023}).
These techniques yield the best results when accompanied by
computational methods that leverage worker-task-label relationships and
their statistical properties.

Many studies in crowdsourcing simplify complex tasks via
multi-classification or post-acceptance steps, as discussed in a pivotal
paper by Bernstein et al. (\protect\hyperlink{ref-Bernstein:10}{2010}).
Meanwhile, researchers in natural language processing and computer
vision develop specialized techniques. However, existing toolkits like
SQUARE (\protect\hyperlink{ref-Sheshadri:13}{Sheshadri \& Lease, 2013}),
CEKA (\protect\hyperlink{ref-Zhang:15}{Zhang et al., 2015}), Truth
Inference (\protect\hyperlink{ref-Zheng:17}{Zheng et al., 2017}),
spark-crowd (\protect\hyperlink{ref-Rodrigo:19}{Rodrigo et al., 2019})
require additional effort for integration into applications, popular
data science libraries and frameworks.

We propose addressing this challenge with \textbf{Crowd-Kit}, an
open-source Python toolkit for computational quality control in
crowdsourcing. Crowd-Kit implements popular quality control methods,
providing a standardized platform for reliable experimentation and
application. We extensively evaluate the Crowd-Kit library to establish
a basis for comparisons. \emph{In all the experiments in this paper, we
used our implementations of the corresponding methods.}

\hypertarget{design}{%
\section{Design}\label{design}}

Our fundamental goal of Crowd-Kit development was to bridge the gap
between crowdsourcing research and vivid data science ecosystem of
NumPy, SciPy, pandas (\protect\hyperlink{ref-McKinney:10}{McKinney,
2010}), and scikit-learn (\protect\hyperlink{ref-Pedregosa:11}{Pedregosa
et al., 2011}). We implemented Crowd-Kit in Python and employed the
highly optimized data structures and algorithms available in these
libraries, maintaining compatibility with the application programming
interface (API) of scikit-learn and data frames/series of pandas. Even
for a user not familiar with crowdsourcing but familiar with scientific
computing and data analysis in Python, the basic API usage would be
straightforward:

\begin{Shaded}
\begin{Highlighting}[]
\CommentTok{\# df is a DataFrame with labeled data in form of (task, label, worker)}
\CommentTok{\# gt is a Series with ground truth per task}
\NormalTok{df, gt }\OperatorTok{=}\NormalTok{ load\_dataset(}\StringTok{\textquotesingle{}relevance{-}2\textquotesingle{}}\NormalTok{)  }\CommentTok{\# binary relevance sample dataset}

\CommentTok{\# run the Dawid{-}Skene categorical aggregation method}
\NormalTok{agg\_ds }\OperatorTok{=}\NormalTok{ DawidSkene(n\_iter}\OperatorTok{=}\DecValTok{10}\NormalTok{).fit\_predict(df)  }\CommentTok{\# same format as gt}
\end{Highlighting}
\end{Shaded}

We implemented all the methods in Crowd-Kit from scratch in Python.
Although unlike spark-crowd (\protect\hyperlink{ref-Rodrigo:19}{Rodrigo
et al., 2019}), our library did not provide a means for running on a
distributed computational cluster, it leveraged efficient
implementations of numerical algorithms in underlying libraries widely
used in the research community. In addition to categorical aggregation
methods, Crowd-Kit offers non-categorical aggregation methods, dataset
loaders, and annotation quality estimators.

\hypertarget{maintenance-and-governance}{%
\section{Maintenance and governance}\label{maintenance-and-governance}}

Crowd-Kit is not bound to any specific crowdsourcing platform, allowing
analyzing data from any crowdsourcing marketplace (as soon as one can
download the labeled data from that platform). Crowd-Kit is an
open-source library working under most operating systems and available
under the Apache license 2.0 both on GitHub and Python Package Index
(PyPI).\footnote{\url{https://github.com/Toloka/crowd-kit} \&
  \url{https://pypi.org/project/crowd-kit/}} All code of Crowd-Kit has
strict type annotations for additional safety and clarity. By the time
of submission, our library had a test coverage of 93\%.

We built Crowd-Kit on top of the established open-source frameworks and
best practices. We widely use the continuous integration facilities via
GitHub Actions for two purposes. First, every patch (\emph{commit} in
git terminology) invokes unit testing and coverage, type checking,
linting, documentation and packaging dry run. Second, every release is
automatically submitted to PyPI directly from GitHub Actions via the
trusted publishing mechanism to avoid potential side effects on the
individual developer machines. Besides commit checks, every code change
(\emph{pull request} on GitHub) goes through a code review by the
Crowd-Kit developers. We accept bug reports via GitHub Issues.

\hypertarget{functionality}{%
\section{Functionality}\label{functionality}}

Crowd-Kit implements a selection of popular methods for answer
aggregation and learning from crowds, dataset loaders, and annotation
quality characteristics.

\hypertarget{aggregating-and-learning-with-crowd-kit}{%
\subsection{Aggregating and learning with
Crowd-Kit}\label{aggregating-and-learning-with-crowd-kit}}

Crowd-Kit features aggregation methods suitable for most kinds of
crowdsourced responses, including categorical, pairwise, sequential, and
image segmentation answers (see the summary in \autoref{tab:methods}).

Methods for \emph{categorical aggregation}, which are the most
widespread in practice, assume that there is only one correct objective
label per task and aim at recovering a latent true label from the
observed noisy data. Some of these methods, such as Dawid-Skene and
GLAD, also estimate latent parameters --- aka skills --- of the workers.
Where the task design does not meet the latent label assumption,
Crowd-Kit offers methods for aggregation \emph{pairwise comparisons},
which are essential for subjective opinion gathering. Also, Crowd-Kit
provides specialized methods for aggregating \emph{sequences} (such as
texts) and \emph{image segmentation}. All these aggregation methods are
implemented purely using NumPy, SciPy, pandas, and scikit-learn without
any deep learning framework. Last but not least, Crowd-Kit offers
methods for \emph{deep learning from crowds} that learn an end-to-end
machine learning model from raw responses submitted by the workers
without the use of aggregation, which are available as ready-to-use
modules for PyTorch (\protect\hyperlink{ref-Paszke:19}{Paszke et al.,
2019}).

One can easily add a new aggregation method to Crowd-Kit. For example,
without the loss of generality, to create a new categorical aggregator,
one should extend the base class \texttt{BaseClassificationAggregator}
and implement two methods, \texttt{fit()} and \texttt{fit\_predict()},
filling the instance variable \texttt{labels\_} with the aggregated
labels.\footnote{See the implementation of Majority Vote at
  \url{https://github.com/Toloka/crowd-kit/blob/main/crowdkit/aggregation/classification/majority_vote.py}
  as an example of an aggregation method.} Also, to add a new method for
learning from crowds, one has to create a subclass from
\texttt{torch.nn.Module} and implement the \texttt{forward()}
method.\footnote{See the implementation of CrowdLayer at
  \url{https://github.com/Toloka/crowd-kit/blob/main/crowdkit/learning/crowd_layer.py}
  as an example of a method for deep learning from crowds.}

\begin{longtable}[]{@{}
  >{\raggedright\arraybackslash}p{(\columnwidth - 2\tabcolsep) * \real{0.2237}}
  >{\raggedright\arraybackslash}p{(\columnwidth - 2\tabcolsep) * \real{0.7763}}@{}}
\caption{Summary of the implemented methods in
Crowd-Kit.\label{tab:methods}}\tabularnewline
\toprule\noalign{}
\begin{minipage}[b]{\linewidth}\raggedright
\textbf{Aggregation}
\end{minipage} & \begin{minipage}[b]{\linewidth}\raggedright
\textbf{Methods}
\end{minipage} \\
\midrule\noalign{}
\endfirsthead
\toprule\noalign{}
\begin{minipage}[b]{\linewidth}\raggedright
\textbf{Aggregation}
\end{minipage} & \begin{minipage}[b]{\linewidth}\raggedright
\textbf{Methods}
\end{minipage} \\
\midrule\noalign{}
\endhead
\bottomrule\noalign{}
\endlastfoot
Categorical & Majority Vote, Wawa (\protect\hyperlink{ref-Wawa}{Appen
Limited, 2021}), Dawid \& Skene
(\protect\hyperlink{ref-Dawid:79}{1979}), \\
& GLAD (\protect\hyperlink{ref-Whitehill:09}{Whitehill et al., 2009}),
MACE (\protect\hyperlink{ref-Hovy:13}{Hovy et al., 2013}), \\
& Karger et al. (\protect\hyperlink{ref-Karger:14}{2014}), M-MSR
(\protect\hyperlink{ref-Ma:20}{Ma \& Olshevsky, 2020}) \\
Pairwise & Bradley \& Terry (\protect\hyperlink{ref-Bradley:52}{1952}),
noisyBT (\protect\hyperlink{ref-Bugakova:19}{Bugakova et al., 2019}) \\
Sequence & ROVER (\protect\hyperlink{ref-Fiscus:97}{Fiscus, 1997}), RASA
and HRRASA (\protect\hyperlink{ref-Li:20}{Li, 2020}), \\
& Language Model
(\protect\hyperlink{ref-Pavlichenko:21:crowdspeech}{Pavlichenko et al.,
2021}) \\
Segmentation & Majority Vote, Expectation-Maximization
(\protect\hyperlink{ref-JungLinLee:18}{Jung-Lin Lee et al., 2018}), \\
& RASA and HRRASA (\protect\hyperlink{ref-Li:20}{Li, 2020}) \\
Learning & CrowdLayer (\protect\hyperlink{ref-Rodrigues:18}{Rodrigues \&
Pereira, 2018}), CoNAL (\protect\hyperlink{ref-Chu:21}{Chu et al.,
2021}) \\
\end{longtable}

\hypertarget{dataset-loaders}{%
\subsection{Dataset loaders}\label{dataset-loaders}}

Crowd-Kit offers convenient dataset loaders for some popular or
demonstrative datasets (see \autoref{tab:datasets}), allowing
downloading them from the Internet in a ready-to-use form with a single
line of code. It is possible to add new datasets in a declarative way
and, if necessary, add the corresponding code to load the data as pandas
data frames and series.

\begin{longtable}[]{@{}
  >{\raggedright\arraybackslash}p{(\columnwidth - 2\tabcolsep) * \real{0.1818}}
  >{\raggedright\arraybackslash}p{(\columnwidth - 2\tabcolsep) * \real{0.8182}}@{}}
\caption{Summary of the datasets provided by
Crowd-Kit.\label{tab:datasets}}\tabularnewline
\toprule\noalign{}
\begin{minipage}[b]{\linewidth}\raggedright
\textbf{Task}
\end{minipage} & \begin{minipage}[b]{\linewidth}\raggedright
\textbf{Datasets}
\end{minipage} \\
\midrule\noalign{}
\endfirsthead
\toprule\noalign{}
\begin{minipage}[b]{\linewidth}\raggedright
\textbf{Task}
\end{minipage} & \begin{minipage}[b]{\linewidth}\raggedright
\textbf{Datasets}
\end{minipage} \\
\midrule\noalign{}
\endhead
\bottomrule\noalign{}
\endlastfoot
Categorical & Toloka Relevance 2 and 5, TREC Relevance
(\protect\hyperlink{ref-Buckley:10}{Buckley et al., 2010}) \\
Pairwise & IMDB-WIKI-SbS
(\protect\hyperlink{ref-Pavlichenko:21:sbs}{Pavlichenko \& Ustalov,
2021}) \\
Sequence & CrowdWSA (\protect\hyperlink{ref-Li:19}{2019}), CrowdSpeech
(\protect\hyperlink{ref-Pavlichenko:21:crowdspeech}{Pavlichenko et al.,
2021}) \\
Image & Common Objects in Context (\protect\hyperlink{ref-Lin:14}{Lin et
al., 2014}) \\
\end{longtable}

\hypertarget{annotation-quality-estimators}{%
\subsection{Annotation quality
estimators}\label{annotation-quality-estimators}}

Crowd-Kit allows one to apply commonly-used techniques to evaluate data
and annotation quality, providing a unified pandas-compatible API to
compute \(\alpha\)
(\protect\hyperlink{ref-Krippendorff:18}{Krippendorff, 2018}),
annotation uncertainty (\protect\hyperlink{ref-Malinin:19}{Malinin,
2019}), agreement with aggregate (\protect\hyperlink{ref-Wawa}{Appen
Limited, 2021}), Dawid-Skene posterior probability, etc.

\hypertarget{evaluation}{%
\section{Evaluation}\label{evaluation}}

We extensively evaluate Crowd-Kit methods for answer aggregation and
learning from crowds. When possible, we compare with other authors;
either way, we show how the currently implemented methods perform on
well-known datasets with noisy crowdsourced data, indicating the
correctness of our implementations.

\hypertarget{evaluation-of-aggregation-methods}{%
\subsection{Evaluation of aggregation
methods}\label{evaluation-of-aggregation-methods}}

\textbf{Categorical.} To ensure the correctness of our implementations,
we compared the observed aggregation quality with the already available
implementations by Zheng et al. (\protect\hyperlink{ref-Zheng:17}{2017})
and Rodrigo et al. (\protect\hyperlink{ref-Rodrigo:19}{2019}).
\autoref{tab:categorical} shows evaluation results, indicating a similar
level of quality as them: \emph{D\_Product}, \emph{D\_PosSent},
\emph{S\_Rel}, and \emph{S\_Adult} are real-world datasets from Zheng et
al. (\protect\hyperlink{ref-Zheng:17}{2017}), and \emph{binary1} and
\emph{binary2} are synthetic datasets from Rodrigo et al.
(\protect\hyperlink{ref-Rodrigo:19}{2019}). Our implementation of M-MSR
could not process the D\_Product dataset in a reasonable time, KOS can
be applied to binary datasets only, and none of our implementations
handled \emph{binary3} and \emph{binary4} synthetic datasets, which
require a distributed computing cluster.

\begin{longtable}[]{@{}
  >{\raggedright\arraybackslash}p{(\columnwidth - 12\tabcolsep) * \real{0.1304}}
  >{\raggedleft\arraybackslash}p{(\columnwidth - 12\tabcolsep) * \real{0.1630}}
  >{\raggedleft\arraybackslash}p{(\columnwidth - 12\tabcolsep) * \real{0.1630}}
  >{\raggedleft\arraybackslash}p{(\columnwidth - 12\tabcolsep) * \real{0.1196}}
  >{\raggedleft\arraybackslash}p{(\columnwidth - 12\tabcolsep) * \real{0.1413}}
  >{\raggedleft\arraybackslash}p{(\columnwidth - 12\tabcolsep) * \real{0.1413}}
  >{\raggedleft\arraybackslash}p{(\columnwidth - 12\tabcolsep) * \real{0.1413}}@{}}
\caption{Comparison of the implemented categorical aggregation methods
(accuracy is used).\label{tab:categorical}}\tabularnewline
\toprule\noalign{}
\begin{minipage}[b]{\linewidth}\raggedright
\textbf{Method}
\end{minipage} & \begin{minipage}[b]{\linewidth}\raggedleft
\textbf{D\_Product}
\end{minipage} & \begin{minipage}[b]{\linewidth}\raggedleft
\textbf{D\_PosSent}
\end{minipage} & \begin{minipage}[b]{\linewidth}\raggedleft
\textbf{S\_Rel}
\end{minipage} & \begin{minipage}[b]{\linewidth}\raggedleft
\textbf{S\_Adult}
\end{minipage} & \begin{minipage}[b]{\linewidth}\raggedleft
\textbf{binary1}
\end{minipage} & \begin{minipage}[b]{\linewidth}\raggedleft
\textbf{binary2}
\end{minipage} \\
\midrule\noalign{}
\endfirsthead
\toprule\noalign{}
\begin{minipage}[b]{\linewidth}\raggedright
\textbf{Method}
\end{minipage} & \begin{minipage}[b]{\linewidth}\raggedleft
\textbf{D\_Product}
\end{minipage} & \begin{minipage}[b]{\linewidth}\raggedleft
\textbf{D\_PosSent}
\end{minipage} & \begin{minipage}[b]{\linewidth}\raggedleft
\textbf{S\_Rel}
\end{minipage} & \begin{minipage}[b]{\linewidth}\raggedleft
\textbf{S\_Adult}
\end{minipage} & \begin{minipage}[b]{\linewidth}\raggedleft
\textbf{binary1}
\end{minipage} & \begin{minipage}[b]{\linewidth}\raggedleft
\textbf{binary2}
\end{minipage} \\
\midrule\noalign{}
\endhead
\bottomrule\noalign{}
\endlastfoot
MV & \(0.897\) & \(0.932\) & \(0.536\) & \(0.763\) & \(0.931\) &
\(0.936\) \\
Wawa & \(0.897\) & \(0.951\) & \(0.557\) & \(0.766\) & \(0.981\) &
\(0.983\) \\
DS & \(0.940\) & \(0.960\) & \(0.615\) & \(0.748\) & \(0.994\) &
\(0.994\) \\
GLAD & \(0.928\) & \(0.948\) & \(0.511\) & \(0.760\) & \(0.994\) &
\(0.994\) \\
KOS & \(0.895\) & \(0.933\) & --- & --- & \(0.993\) & \(0.994\) \\
MACE & \(0.929\) & \(0.950\) & \(0.501\) & \(0.763\) & \(0.995\) &
\(0.995\) \\
M-MSR & --- & \(0.937\) & \(0.425\) & \(0.751\) & \(0.994\) &
\(0.994\) \\
\end{longtable}

\textbf{Pairwise.} \autoref{tab:pairwise} shows the comparison of the
\emph{Bradley-Terry} and \emph{noisyBT} methods implemented in Crowd-Kit
to the random baseline on the graded readability dataset by Chen et al.
(\protect\hyperlink{ref-Chen:13}{2013}) and a larger people age dataset
by Pavlichenko \& Ustalov
(\protect\hyperlink{ref-Pavlichenko:21:sbs}{2021}).

\begin{longtable}[]{@{}lrr@{}}
\caption{Comparison of implemented pairwise aggregation methods
(Spearman's \(\rho\) is used).\label{tab:pairwise}}\tabularnewline
\toprule\noalign{}
\textbf{Method} & \textbf{Chen et al.
(\protect\hyperlink{ref-Chen:13}{2013})} & \textbf{IMDB-WIKI-SBS} \\
\midrule\noalign{}
\endfirsthead
\toprule\noalign{}
\textbf{Method} & \textbf{Chen et al.
(\protect\hyperlink{ref-Chen:13}{2013})} & \textbf{IMDB-WIKI-SBS} \\
\midrule\noalign{}
\endhead
\bottomrule\noalign{}
\endlastfoot
Bradley-Terry & \(0.246\) & \(0.737\) \\
noisyBT & \(0.238\) & \(0.744\) \\
Random & \(-0.013\) & \(-0.001\) \\
\end{longtable}

\textbf{Sequence.} We used two datasets, CrowdWSA
(\protect\hyperlink{ref-Li:19}{Li \& Fukumoto, 2019}) and CrowdSpeech
(\protect\hyperlink{ref-Pavlichenko:21:crowdspeech}{Pavlichenko et al.,
2021}). As the typical application for sequence aggregation in
crowdsourcing is audio transcription, we used the word error rate as the
quality criterion (\protect\hyperlink{ref-Fiscus:97}{Fiscus, 1997}) in
\autoref{tab:sequence}.

\begin{longtable}[]{@{}lcrrr@{}}
\caption{Comparison of implemented sequence aggregation methods (average
word error rate is used).\label{tab:sequence}}\tabularnewline
\toprule\noalign{}
\textbf{Dataset} & \textbf{Version} & \textbf{ROVER} & \textbf{RASA} &
\textbf{HRRASA} \\
\midrule\noalign{}
\endfirsthead
\toprule\noalign{}
\textbf{Dataset} & \textbf{Version} & \textbf{ROVER} & \textbf{RASA} &
\textbf{HRRASA} \\
\midrule\noalign{}
\endhead
\bottomrule\noalign{}
\endlastfoot
CrowdWSA & J1 & \(0.612\) & \(0.659\) & \(0.676\) \\
& T1 & \(0.514\) & \(0.483\) & \(0.500\) \\
& T2 & \(0.524\) & \(0.498\) & \(0.520\) \\
CrowdSpeech & dev-clean & \(0.676\) & \(0.750\) & \(0.745\) \\
& dev-other & \(0.132\) & \(0.142\) & \(0.142\) \\
& test-clean & \(0.729\) & \(0.860\) & \(0.859\) \\
& test-other & \(0.134\) & \(0.157\) & \(0.157\) \\
\end{longtable}

\textbf{Segmentation.} We annotated on the Toloka crowdsourcing platform
a sample of 2,000 images from the MS COCO
(\protect\hyperlink{ref-Lin:14}{Lin et al., 2014}) dataset consisting of
four object labels. For each image, nine workers submitted
segmentations. The dataset is available in Crowd-Kit as
\texttt{mscoco\_small}. In total, we received 18,000 responses.
\autoref{tab:segmentation} shows the comparison of the methods on the
above-described dataset using the \emph{intersection over union} (IoU)
criterion.

\begin{longtable}[]{@{}lrrr@{}}
\caption{Comparison of implemented image aggregation algorithms (IoU is
used).\label{tab:segmentation}}\tabularnewline
\toprule\noalign{}
\textbf{Dataset} & \textbf{MV} & \textbf{EM} & \textbf{RASA} \\
\midrule\noalign{}
\endfirsthead
\toprule\noalign{}
\textbf{Dataset} & \textbf{MV} & \textbf{EM} & \textbf{RASA} \\
\midrule\noalign{}
\endhead
\bottomrule\noalign{}
\endlastfoot
MS COCO & \(0.839\) & \(0.861\) & \(0.849\) \\
\end{longtable}

\hypertarget{evaluation-of-methods-for-learning-from-crowds}{%
\subsection{Evaluation of methods for learning from
crowds}\label{evaluation-of-methods-for-learning-from-crowds}}

To demonstrate the impact of learning on raw annotator labels compared
to answer aggregation in crowdsourcing, we compared the implemented
methods for learning from crowds with the two classical aggregation
algorithms, Majority Vote (MV) and Dawid-Skene (DS). We picked the two
most common machine learning tasks for which ground truth datasets are
available: text classification and image classification. For text
classification, we used the IMDB Movie Reviews dataset
(\protect\hyperlink{ref-Maas:11}{Maas et al., 2011}), and for image
classification, we chose CIFAR-10
(\protect\hyperlink{ref-Krizhevsky:09}{Krizhevsky, 2009}). In each
dataset, each object was annotated by three different annotators; 100
objects were used as golden tasks.

We compared how different methods for learning from crowds impact test
accuracy. We picked two different backbone networks for text
classification, LSTM (\protect\hyperlink{ref-Hochreiter:97}{Hochreiter
\& Schmidhuber, 1997}) and RoBERTa (\protect\hyperlink{ref-Liu:19}{Liu
et al., 2019}), and one backbone network for image classification,
VGG-16 (\protect\hyperlink{ref-Simonyan:15}{Simonyan \& Zisserman,
2015}). Then, we trained each backbone in three scenarios: use the fully
connected layer after the backbone without taking into account any
specifics of crowdsourcing (Base), CrowdLayer method by Rodrigues \&
Pereira (\protect\hyperlink{ref-Rodrigues:18}{2018}), and CoNAL method
by Chu et al. (\protect\hyperlink{ref-Chu:21}{2021}).
\autoref{tab:learning} shows the evaluation results.

\begin{longtable}[]{@{}
  >{\raggedright\arraybackslash}p{(\columnwidth - 12\tabcolsep) * \real{0.1687}}
  >{\centering\arraybackslash}p{(\columnwidth - 12\tabcolsep) * \real{0.1687}}
  >{\raggedleft\arraybackslash}p{(\columnwidth - 12\tabcolsep) * \real{0.1325}}
  >{\raggedleft\arraybackslash}p{(\columnwidth - 12\tabcolsep) * \real{0.1928}}
  >{\raggedleft\arraybackslash}p{(\columnwidth - 12\tabcolsep) * \real{0.1205}}
  >{\raggedleft\arraybackslash}p{(\columnwidth - 12\tabcolsep) * \real{0.1084}}
  >{\raggedleft\arraybackslash}p{(\columnwidth - 12\tabcolsep) * \real{0.1084}}@{}}
\caption{Comparison of different methods for deep learning from crowds
with traditional answer aggregation methods (test set accuracy is
used).\label{tab:learning}}\tabularnewline
\toprule\noalign{}
\begin{minipage}[b]{\linewidth}\raggedright
\textbf{Dataset}
\end{minipage} & \begin{minipage}[b]{\linewidth}\centering
\textbf{Backbone}
\end{minipage} & \begin{minipage}[b]{\linewidth}\raggedleft
\textbf{CoNAL}
\end{minipage} & \begin{minipage}[b]{\linewidth}\raggedleft
\textbf{CrowdLayer}
\end{minipage} & \begin{minipage}[b]{\linewidth}\raggedleft
\textbf{Base}
\end{minipage} & \begin{minipage}[b]{\linewidth}\raggedleft
\textbf{DS}
\end{minipage} & \begin{minipage}[b]{\linewidth}\raggedleft
\textbf{MV}
\end{minipage} \\
\midrule\noalign{}
\endfirsthead
\toprule\noalign{}
\begin{minipage}[b]{\linewidth}\raggedright
\textbf{Dataset}
\end{minipage} & \begin{minipage}[b]{\linewidth}\centering
\textbf{Backbone}
\end{minipage} & \begin{minipage}[b]{\linewidth}\raggedleft
\textbf{CoNAL}
\end{minipage} & \begin{minipage}[b]{\linewidth}\raggedleft
\textbf{CrowdLayer}
\end{minipage} & \begin{minipage}[b]{\linewidth}\raggedleft
\textbf{Base}
\end{minipage} & \begin{minipage}[b]{\linewidth}\raggedleft
\textbf{DS}
\end{minipage} & \begin{minipage}[b]{\linewidth}\raggedleft
\textbf{MV}
\end{minipage} \\
\midrule\noalign{}
\endhead
\bottomrule\noalign{}
\endlastfoot
IMDb & LSTM & \(0.844\) & \(0.825\) & \(0.835\) & \(0.841\) &
\(0.819\) \\
IMDb & RoBERTa & \(0.932\) & \(0.928\) & \(0.927\) & \(0.932\) &
\(0.927\) \\
CIFAR-10 & VGG-16 & \(0.825\) & \(0.863\) & \(0.882\) & \(0.877\) &
\(0.865\) \\
\end{longtable}

Our experiment shows the feasibility of training a deep learning model
directly from the raw annotated data, skipping trivial aggregation
methods like MV. However, specialized methods like CoNAL and CrowdLayer
or non-trivial aggregation methods like DS can significantly enhance
prediction accuracy. It is crucial to make a well-informed model
selection to achieve optimal results. We believe that Crowd-Kit can
seamlessly integrate these methods into machine learning pipelines that
utilize crowdsourced data with reliability and ease.

\hypertarget{conclusion}{%
\section{Conclusion}\label{conclusion}}

Our experience running Crowd-Kit in production for processing
crowdsourced data at Toloka shows that it successfully handles
industry-scale datasets without needing a large compute cluster. We
believe that the availability of computational quality control
techniques in a standardized way would open new venues for reliable
improvement of the crowdsourcing quality beyond the traditional
well-known methods and pipelines.

\hypertarget{acknowledgements}{%
\section{Acknowledgements}\label{acknowledgements}}

The work was done while the authors were with Yandex. We are grateful to
Enrique G. Rodrigo for sharing the spark-crowd evaluation dataset. We
want to thank Daniil Fedulov, Iulian Giliazev, Artem Grigorev, Daniil
Likhobaba, Vladimir Losev, Stepan Nosov, Alisa Smirnova, Aleksey
Sukhorosov, and Evgeny Tulin for their contributions to the library.
Last but not least, we appreciate the improvements to our library made
by open-source
\href{https://github.com/Toloka/crowd-kit/graphs/contributors}{contributors}
and the reviewers of this paper. We received no external funding.

\hypertarget{references}{%
\section*{References}\label{references}}
\addcontentsline{toc}{section}{References}

\hypertarget{refs}{}
\begin{CSLReferences}{1}{0}
\leavevmode\vadjust pre{\hypertarget{ref-Wawa}{}}%
Appen Limited. (2021). \emph{{Calculating Worker Agreement with
Aggregate (Wawa)}}.
\url{https://success.appen.com/hc/en-us/articles/202703205-Calculating-Worker-Agreement-with-Aggregate-Wawa-}

\leavevmode\vadjust pre{\hypertarget{ref-Bernstein:10}{}}%
Bernstein, M. S., Little, G., Miller, R. C., Hartmann, B., Ackerman, M.
S., Karger, D. R., Crowell, D., \& Panovich, K. (2010). {Soylent: A Word
Processor with a Crowd Inside}. \emph{Proceedings of the 23Nd Annual ACM
Symposium on User Interface Software and Technology}, 313--322.
\url{https://doi.org/10.1145/1866029.1866078}

\leavevmode\vadjust pre{\hypertarget{ref-Bradley:52}{}}%
Bradley, R. A., \& Terry, M. E. (1952). {Rank Analysis of Incomplete
Block Designs: I. The Method of Paired Comparisons}. \emph{Biometrika},
\emph{39}(3/4), 324--345. \url{https://doi.org/10.2307/2334029}

\leavevmode\vadjust pre{\hypertarget{ref-Buckley:10}{}}%
Buckley, C., Lease, M., \& Smucker, M. D. (2010). {Overview of the TREC
2010 Relevance Feedback Track (Notebook)}. \emph{The Nineteenth TREC
Notebook}.
\url{https://www.ischool.utexas.edu/~ml/papers/trec-notebook-2010.pdf}

\leavevmode\vadjust pre{\hypertarget{ref-Bugakova:19}{}}%
Bugakova, N., Fedorova, V., Gusev, G., \& Drutsa, A. (2019).
{Aggregation of pairwise comparisons with reduction of biases}.
\emph{2019 ICML Workshop on Human in the Loop Learning}.
\url{https://arxiv.org/abs/1906.03711}

\leavevmode\vadjust pre{\hypertarget{ref-Chen:13}{}}%
Chen, X., Bennett, P. N., Collins-Thompson, K., \& Horvitz, E. (2013).
{Pairwise Ranking Aggregation in a Crowdsourced Setting}.
\emph{Proceedings of the Sixth ACM International Conference on Web
Search and Data Mining}, 193--202.
\url{https://doi.org/10.1145/2433396.2433420}

\leavevmode\vadjust pre{\hypertarget{ref-Chu:21}{}}%
Chu, Z., Ma, J., \& Wang, H. (2021). {Learning from Crowds by Modeling
Common Confusions}. \emph{Proceedings of the AAAI Conference on
Artificial Intelligence}, \emph{35}(7), 5832--5840.
\url{https://doi.org/10.1609/aaai.v35i7.16730}

\leavevmode\vadjust pre{\hypertarget{ref-Dawid:79}{}}%
Dawid, A. P., \& Skene, A. M. (1979). {Maximum Likelihood Estimation of
Observer Error-Rates Using the EM Algorithm}. \emph{Journal of the Royal
Statistical Society, Series~C (Applied Statistics)}, \emph{28}(1),
20--28. \url{https://doi.org/10.2307/2346806}

\leavevmode\vadjust pre{\hypertarget{ref-Fiscus:97}{}}%
Fiscus, J. G. (1997). {A post-processing system to yield reduced word
error rates: Recognizer Output Voting Error Reduction (ROVER)}.
\emph{1997 IEEE Workshop on Automatic Speech Recognition and
Understanding Proceedings}, 347--354.
\url{https://doi.org/10.1109/ASRU.1997.659110}

\leavevmode\vadjust pre{\hypertarget{ref-Hochreiter:97}{}}%
Hochreiter, S., \& Schmidhuber, J. (1997). {Long Short-Term Memory}.
\emph{Neural Computation}, \emph{9}(8), 1735--1780.
\url{https://doi.org/10.1162/neco.1997.9.8.1735}

\leavevmode\vadjust pre{\hypertarget{ref-Hovy:13}{}}%
Hovy, D., Berg-Kirkpatrick, T., Vaswani, A., \& Hovy, E. (2013).
{Learning Whom to Trust with MACE}. \emph{Proceedings of the 2013
Conference of the North American Chapter of the Association for
Computational Linguistics: Human Language Technologies}, 1120--1130.
\url{https://aclanthology.org/N13-1132}

\leavevmode\vadjust pre{\hypertarget{ref-JungLinLee:18}{}}%
Jung-Lin Lee, D., Das Sarma, A., \& Parameswaran, A. (2018).
\emph{{Quality Evaluation Methods for Crowdsourced Image Segmentation}}
{[}Technical Report{]}. Stanford University; Stanford InfoLab.
\url{http://ilpubs.stanford.edu:8090/1161/}

\leavevmode\vadjust pre{\hypertarget{ref-Karger:14}{}}%
Karger, D. R., Oh, S., \& Shah, D. (2014). {Budget-Optimal Task
Allocation for Reliable Crowdsourcing Systems}. \emph{Operations
Research}, \emph{62}(1), 1--24.
\url{https://doi.org/10.1287/opre.2013.1235}

\leavevmode\vadjust pre{\hypertarget{ref-Krippendorff:18}{}}%
Krippendorff, K. (2018). \emph{{Content Analysis: An Introduction to Its
Methodology}} (Fourth Edition). SAGE Publications, Inc.
ISBN:~978-1-5063-9566-1

\leavevmode\vadjust pre{\hypertarget{ref-Krizhevsky:09}{}}%
Krizhevsky, A. (2009). \emph{{Learning Multiple Layers of Features from
Tiny Images}}. University of Toronto.
\url{https://www.cs.toronto.edu/~kriz/learning-features-2009-TR.pdf}

\leavevmode\vadjust pre{\hypertarget{ref-Li:20}{}}%
Li, J. (2020). {Crowdsourced Text Sequence Aggregation Based on Hybrid
Reliability and Representation}. \emph{Proceedings of the 43rd
International ACM SIGIR Conference on Research and Development in
Information Retrieval}, 1761--1764.
\url{https://doi.org/10.1145/3397271.3401239}

\leavevmode\vadjust pre{\hypertarget{ref-Li:19}{}}%
Li, J., \& Fukumoto, F. (2019). {A Dataset of Crowdsourced Word
Sequences: Collections and Answer Aggregation for Ground Truth
Creation}. \emph{Proceedings of the First Workshop on Aggregating and
Analysing Crowdsourced Annotations for NLP}, 24--28.
\url{https://doi.org/10.18653/v1/D19-5904}

\leavevmode\vadjust pre{\hypertarget{ref-Lin:14}{}}%
Lin, T.-Y., Maire, M., Belongie, S., Hays, J., Perona, P., Ramanan, D.,
Dollár, P., \& Zitnick, C. L. (2014). {Microsoft COCO: Common Objects in
Context}. \emph{Computer Vision -- ECCV 2014}, 740--755.
\url{https://doi.org/10.1007/978-3-319-10602-1_48}

\leavevmode\vadjust pre{\hypertarget{ref-Liu:19}{}}%
Liu, Y., Ott, M., Goyal, N., Du, J., Joshi, M., Chen, D., Levy, O.,
Lewis, M., Zettlemoyer, L., \& Stoyanov, V. (2019). \emph{{RoBERTa: A
Robustly Optimized BERT Pretraining Approach}}.
\url{https://arxiv.org/abs/1907.11692}

\leavevmode\vadjust pre{\hypertarget{ref-Ma:20}{}}%
Ma, Q., \& Olshevsky, A. (2020). {Adversarial Crowdsourcing Through
Robust Rank-One Matrix Completion}. \emph{Advances in Neural Information
Processing Systems~33}, 21841--21852.
\url{https://proceedings.neurips.cc/paper/2020/file/f86890095c957e9b949d11d15f0d0cd5-Paper.pdf}

\leavevmode\vadjust pre{\hypertarget{ref-Maas:11}{}}%
Maas, A. L., Daly, R. E., Pham, P. T., Huang, D., Ng, A. Y., \& Potts,
C. (2011). {Learning Word Vectors for Sentiment Analysis}.
\emph{Proceedings of the 49th Annual Meeting of the Association for
Computational Linguistics: Human Language Technologies}, 142--150.
\url{https://aclanthology.org/P11-1015}

\leavevmode\vadjust pre{\hypertarget{ref-Malinin:19}{}}%
Malinin, A. (2019). \emph{{Uncertainty Estimation in Deep Learning with
application to Spoken Language Assessment}} {[}PhD thesis, University of
Cambridge{]}. \url{https://doi.org/10.17863/CAM.45912}

\leavevmode\vadjust pre{\hypertarget{ref-McKinney:10}{}}%
McKinney, W. (2010). {Data Structures for Statistical Computing in
Python}. \emph{Proceedings of the 9th Python in Science Conference},
56--61. \url{https://doi.org/10.25080/Majora-92bf1922-00a}

\leavevmode\vadjust pre{\hypertarget{ref-Paszke:19}{}}%
Paszke, A., Gross, S., Massa, F., Lerer, A., Bradbury, J., Chanan, G.,
Killeen, T., Lin, Z., Gimelshein, N., Antiga, L., Desmaison, A., Kopf,
A., Yang, E., DeVito, Z., Raison, M., Tejani, A., Chilamkurthy, S.,
Steiner, B., Fang, L., \ldots{} Chintala, S. (2019). {PyTorch: An
Imperative Style, High-Performance Deep Learning Library}.
\emph{Advances in Neural Information Processing Systems}, \emph{32}.
\url{https://proceedings.neurips.cc/paper/2019/file/bdbca288fee7f92f2bfa9f7012727740-Paper.pdf}

\leavevmode\vadjust pre{\hypertarget{ref-Pavlichenko:21:crowdspeech}{}}%
Pavlichenko, N., Stelmakh, I., \& Ustalov, D. (2021). {CrowdSpeech and
Vox~DIY: Benchmark Dataset for Crowdsourced Audio Transcription}.
\emph{Proceedings of the Neural Information Processing Systems Track on
Datasets and Benchmarks}.
\url{https://datasets-benchmarks-proceedings.neurips.cc/paper/2021/hash/65ded5353c5ee48d0b7d48c591b8f430-Abstract-round1.html}

\leavevmode\vadjust pre{\hypertarget{ref-Pavlichenko:21:sbs}{}}%
Pavlichenko, N., \& Ustalov, D. (2021). {IMDB-WIKI-SbS: An Evaluation
Dataset for Crowdsourced Pairwise Comparisons}. \emph{NeurIPS
Data-Centric AI Workshop}.
\url{https://datacentricai.org/neurips21/papers/115_CameraReady_NeurIPS_2021_Data_Centric_AI_IMDB_WIKI_SbS-2.pdf}

\leavevmode\vadjust pre{\hypertarget{ref-Pedregosa:11}{}}%
Pedregosa, F., Varoquaux, G., Gramfort, A., Michel, V., Thirion, B.,
Grisel, O., Blondel, M., Prettenhofer, P., Weiss, R., Dubourg, V.,
Vanderplas, J., Passos, A., Cournapeau, D., Brucher, M., Perrot, M., \&
Duchesnay, É. (2011). {Scikit-learn: Machine Learning in Python}.
\emph{Journal of Machine Learning Research}, \emph{12}(85), 2825--2830.
\url{https://jmlr.org/papers/v12/pedregosa11a.html}

\leavevmode\vadjust pre{\hypertarget{ref-Rodrigo:19}{}}%
Rodrigo, E. G., Aledo, J. A., \& Gámez, J. A. (2019). {spark-crowd: A
Spark Package for Learning from Crowdsourced Big Data}. \emph{Journal of
Machine Learning Research}, \emph{20}, 1--5.
\url{https://jmlr.org/papers/v20/17-743.html}

\leavevmode\vadjust pre{\hypertarget{ref-Rodrigues:18}{}}%
Rodrigues, F., \& Pereira, F. C. (2018). {Deep Learning from Crowds}.
\emph{Proceedings of the AAAI Conference on Artificial Intelligence},
\emph{32}(1), 1611--1618. \url{https://doi.org/10.1609/aaai.v32i1.11506}

\leavevmode\vadjust pre{\hypertarget{ref-Sheshadri:13}{}}%
Sheshadri, A., \& Lease, M. (2013). {SQUARE: A Benchmark for Research on
Computing Crowd Consensus}. \emph{Proceedings of the AAAI Conference on
Human Computation and Crowdsourcing}, \emph{1}(1), 156--164.
\url{https://doi.org/10.1609/hcomp.v1i1.13088}

\leavevmode\vadjust pre{\hypertarget{ref-Simonyan:15}{}}%
Simonyan, K., \& Zisserman, A. (2015). \emph{{Very Deep Convolutional
Networks for Large-Scale Image Recognition}}.
\url{https://arxiv.org/abs/1409.1556}

\leavevmode\vadjust pre{\hypertarget{ref-Whitehill:09}{}}%
Whitehill, J., Wu, T., Bergsma, J., Movellan, J. R., \& Ruvolo, P. L.
(2009).
\href{https://papers.nips.cc/paper/3644-whose-vote-should-count-more-optimal-integration-of-labels-from-labelers-of-unknown-expertise.pdf}{{Whose
Vote Should Count More: Optimal Integration of Labels from Labelers of
Unknown Expertise}}. In \emph{Advances in neural information processing
systems 22} (pp. 2035--2043). Curran Associates, Inc.
ISBN:~978-1-61567-911-9

\leavevmode\vadjust pre{\hypertarget{ref-Zhang:15}{}}%
Zhang, J., Sheng, V. S., Nicholson, B. A., \& Wu, X. (2015). {CEKA: A
Tool for Mining the Wisdom of Crowds}. \emph{Journal of Machine Learning
Research}, \emph{16}(88), 2853--2858.
\url{https://jmlr.org/papers/v16/zhang15a.html}

\leavevmode\vadjust pre{\hypertarget{ref-Zhdanovskaya:23}{}}%
Zhdanovskaya, A., Baidakova, D., \& Ustalov, D. (2023). {Data Labeling
for Machine Learning Engineers: Project-Based Curriculum and
Data-Centric Competitions}. \emph{Proceedings of the AAAI Conference on
Artificial Intelligence}, \emph{37}(13), 15886--15893.
\url{https://doi.org/10.1609/aaai.v37i13.26886}

\leavevmode\vadjust pre{\hypertarget{ref-Zheng:17}{}}%
Zheng, Y., Li, G., Li, Y., Shan, C., \& Cheng, R. (2017). {Truth
Inference in Crowdsourcing: Is the Problem Solved?} \emph{Proceedings of
the VLDB Endowment}, \emph{10}(5), 541--552.
\url{https://doi.org/10.14778/3055540.3055547}

\end{CSLReferences}

\end{document}